\documentclass[twocolumn,showpacs,preprintnumbers,amsmath,amssymb,10pt,prd]{revtex4}

\usepackage{graphicx}
\usepackage{dcolumn}
\usepackage{bm}
\usepackage{hyperref}
\usepackage{color}
\usepackage{mathrsfs}

\newcommand\be{\begin{equation}}
\newcommand\ee{\end{equation}}
\newcommand\ba{\begin{eqnarray}}
\newcommand\ea{\end{eqnarray}}

\newcommand\ham{{\cal H}}
\newcommand\mub{\bar{\mu}}
\newcommand\dotp{\dot{p}}

\newcommand\doth{\dot{h}}
\newcommand\sQ{{\sigma_{Q}^2}}
\newcommand\cl{{(0)}}

\begin{document}
\title{Modified Friedmann equation and survey of solutions in effective Bianchi-I loop quantum cosmology}

\author{Linda Linsefors}%
 \email{linsefors@lpsc.in2p3.fr}
\affiliation{%
Laboratoire de Physique Subatomique et de Cosmologie, UJF, INPG, CNRS, IN2P3\\
53, avenue des Martyrs, 38026 Grenoble cedex, France
}%

\author{Aurelien Barrau}%
 \email{Aurelien.Barrau@cern.ch}
\affiliation{%
Laboratoire de Physique Subatomique et de Cosmologie, UJF, INPG, CNRS, IN2P3\\
53,avenue des Martyrs, 38026 Grenoble cedex, France
}%

\date{\today}

\begin{abstract}
In this article, we study the equations driving the dynamics of a Bianchi-I universe described by holonomy corrected effective loop quantum cosmology. We derive the LQC-modified generalized Friedmann equation, which is used as a guide to find different types of solutions. It turns out that, in this framework, most solutions never reach the classical behavior.
\end{abstract}


\maketitle

\section{Introduction}

Loop quantum gravity (LQG) is a tentative nonperturbative and background-independent quantization
of general relativity. It uses Ashtekar variables, namely SU(2) valued connections and conjugate densitized triads. The quantization is obtained through holonomies of the connections and fluxes of the densitized triads (see, 
{\it e.g.}, \cite{rovelli1} for introductions). Basically, loop quantum cosmology (LQC)
is the symmetry reduced version of LQG. In LQC, the big bang is generically replaced by a big bounce due to huge repulsive quantum geometrical effects (see, {\it e.g.}, \cite{lqc_review} for reviews).\\

In bouncing cosmologies, the issue of anisotropies is however crucial for a simple reason: the shear term basically scales as $1/a^6$ where $a$ is the scale factor of the Universe. Therefore, when the Universe is in its contraction phase, it is expected that the shear term eventually dominates and drives the dynamics. When spatial homogeneity is assumed, anisotropic hypersurfaces admit transitive groups of motion that must be three- or four-parameters isometry groups. The four-parameters groups admitting no simply transitive subgroups will not be considered here. There are nine algebraically inequivalent three-parameters simply transitive Lie groups, denoted Bianchi I through IX, with well known structure constants. The flat, closed and open generalizations of the FLRW model are respectively Bianchi-I, Bianchi-IX and Bianchi-V. As the Universe is nearly flat today and as the relative weight of the curvature term in the Friedmann equation is decreasing with decreasing values of the scale factor, it is reasonable to focus on the Bianchi-I model to study the dynamics around the bounce.\\

Many studies have already been devoted to Bianchi-I LQC \cite{CV,mubar,bianchiI_lqc}. In particular, it was shown that the bounce prediction is robust. As the main features of isotropic LQC are well captured by semi-classical effective equations, and it is a good guess that this remains true in the extended Bianchi-I case. The solutions of effective equations were studied into the details in \cite{Gupt:2013swa}. In this work, we focus on slightly different aspects and derive the LQC-modified generalized Friedmann equation that was still missing. Thanks to this equation, we have systematically explored the full solution space in a way that hasn't been tired before.

\section{Classical equations}

The metric for a Bianchi-I spacetime reads as:
\be
ds := -N^2 d\tau^2 + a_1^2dx^2 + a_2^2dy^2 + a_3^2dz^2,
\ee
where $a_i$ denote the directional scale factors. A dot means derivation with  respect to the cosmic time $t$, with $dt=Nd\tau$.

Classically, the evolution of this metric is described by the Hamiltonian 
\be
\ham = \ham_G(c_i,p_i) + \ham_M(p_i,\phi_n,\pi_n),
\label{classham}
\ee
where
\be
\ham_G = \frac{N}{\kappa\gamma^2}\left(\sqrt{\frac{p_1 p_2}{p_3}}c_1c_2+\sqrt{\frac{p_2 p_3}{p_1}}c_2c_2+\sqrt{\frac{p_3 p_1}{p_2}}c_2c_3\right),
\label{classhamG}
\ee
and
\be
\ham_M = N \sqrt{p_1p_2p_3}\ \rho,
\ee
with the Poison brackets
\be
\{c_i,p_j\}=\kappa\gamma\delta_{ij}\quad ,\quad \{\phi_n,\pi_m\}=\delta_{mn},
\ee
where $i,j\in\{1,2,3\}$ and $n,m\in\{1,2,\dots,M\}$ for $M$ matter fields. In the following, we have chosen to consider a comoving volume of size $1\times1\times1$. Since the universe is assumed to be homogenous, this will not affect the results. We denote by $\phi_n$ the matter fields, $\pi_n$ their conjugate momentum, and $\rho$ the total matter density. The $c_i$ and $p_i$ entering Eq. (\ref{classhamG}) are the diagonal elements of the Ashtekar variables ($p_i$ is assumed to always be positive).

The directional scale factors can be written as
\be
a_1 = \sqrt{\frac{p_2 p_3}{p_1}}
\qquad \text{and cyclic expressions.}
\ee
The generalized Friedmann equation is
\be
H^2=\sigma^2+\frac{\kappa}{3}\rho,
\label{clfried}
\ee
where
\be
H:=\frac{\dot{a}}{a}=\frac{1}{3}(H_1+H_2+H_3),
\ee
\be
a:=(a_1a_2a_3)^{1/3},
\ee
\be
H_1:=\frac{\dot{a}_1}{a_1}=
-\frac{\dot{p}_1}{2p_1}+\frac{\dot{p}_3}{2p_3}+\frac{\dot{p}_3}{2p_3}
\qquad \text{and cyclic expressions,}
\ee
\be
\sigma^2:=\frac{1}{18}\Big[ (H_1-H_2)^2+(H_2-H_3)^2+(H_3-H_1)^2 \Big].
\label{sigma}
\ee
It should be pointed out that the $1/18$ factor is not used in similar studies.

If we assume isotropic matter, that is 
\be
\ham_M(p_i,\phi,\pi)=\ham_M(\sqrt{p_1p_2p_3},\phi,\pi),
\label{isom}
\ee 
then the equations of motion for $H_i$ become
\be
\dot{H}_1=-H_1^2+H_2H_3-\frac{\kappa}{2}(\rho+P) 
\qquad \text{and cyclic terms,}
\label{cleom}
\ee
where $P$ is defined to fulfill the equation $\dot{\rho}=3H(\rho+P)$, that is
\be
P:=-\frac{\partial(\ham_m/N)}{\partial\sqrt{p_1p_2p_3}}.
\ee

Several other relations will be useful:
\be
\dot{H}_i-\dot{H}_j=-3H(H_i-H_j) \quad\Leftrightarrow\quad H_i-H_j\propto a^{-3},
\label{interi}
\ee
leading to
\be
\sigma^2 \propto a^{-6} \quad\text{and}\quad \frac{H_i-H_j}{H_i-H_k}=\text{constant}.
\label{interf}
\ee
Classically $H_i$ can change sign, but $H$ cannot. Many details about the classical behaviors of a Bianchi-I universe
can be found, {\it e.g.}, in \cite{phd}.\\ \\

\section{Effective holonomy corrections}

The holonomy correction in effective LQC is due to the fact that the Ashtekar connection cannot be promoted to be an operator but only its holonomy can. It is believed to capture most quantum effects at the semi-classical level. Following the usual prescription, we perform the substitution
\be
c_i\rightarrow \frac{\sin(\mub_i c_i)}{\mub_i}
\ee
in the Hamiltonian given by Eqs (\ref{classham}) and (\ref{classhamG}). The $\mub_i$ are given by
\be
\mub_1 = \lambda\sqrt{\frac{p_1}{p_2p_3}}
\qquad \text{and cyclic expressions,}
\ee
where $\lambda$ is the square root of the minimum area eigenvalue of the LQG area operator ($\lambda=\sqrt{\Delta}$). This was first proposed in \cite{CV}, and later derived in \cite{mubar}.

The effective holonomy corrected gravitational Hamiltonian is
\begin{widetext}
\be
\ham_G=-\frac{N\sqrt{p_1p_2p_3}}{\kappa\ \gamma^2\lambda^2}\Big[\sin(\mub_1c_1)\sin(\mub_2c_2)+\sin(\mub_2c_2)\sin(\mub_3c_3)+\sin(\mub_3c_3)\sin(\mub_1c_1)\Big].
\ee
\end{widetext}
The matter Hamiltonian $\ham_M$ remains unchanged.

\section{The LQC-modified generalized Friedmann equation}
\label{LQC}

Various versions of the Friedmann equation --depending on the specific model considered-- are used in cosmology. They allow to derive the key features of the dynamics in a simple way. The LQC-modified generalized Friedmann equation describing a holonomy-corrected Bianchi-I universe has so far been missing. It is derived in this section and, in more details, in Appendix \ref{app}.

The Friedmann equation is found by rewriting the constraint $\ham=0$ in therms of physical parameters. In our case, these parameters are: the total Hubble parameter, matter density and shear. We start by finding the directional and total Hubble parameters as functions of $c_i$ and $p_i$:

\begin{multline}
\dotp_1=\frac{1}{N}\{p_1,\ham\}=\frac{p_1}{\gamma\lambda}\cos(\mub_1c_1)\Big[\sin(\mub_2c_2)+\sin(\mub_3c_3)\Big]\\
\text{and cyclic expressions.}
\end{multline}
From this, we get the directional Hubble parameters $H_i$ and total Hubble parameter $H$:
\begin{widetext}
\be
H_1=-\frac{\dot{p}_1}{2p_1}+\frac{\dot{p}_2}{2p_2}+\frac{\dot{p}_3}{2p_3}=\frac{1}{2\gamma\lambda}\Big[\sin(\mub_2c_2+\mub_3c_3)+\sin(\mub_1c_1-\mub_2c_2)+\sin(\mub_1c_1-\mub_3c_3)\Big]
\qquad \text{and cyclic,}
\label{h}
\ee
\be
H:=\frac{1}{3}(H_1+H_2+H_3)=\frac{1}{6\gamma\lambda}\Big[\sin(\mub_1c_1+\mub_2c_2)+\sin(\mub_2c_2+\mub_3c_3)+\sin(\mub_3c_3+\mub_1c_1)\Big].
\label{H}
\ee
We also define the "quantum shear" as:
\be
\sQ:=\frac{1}{3\lambda^2\gamma^2}\left(1-\frac{1}{3}\Big[\cos(\mub_1c_1-\mub_2c_2)+\cos(\mub_2c_2-\mub_3c_3)+\cos(\mub_3c_3-\mub_1c_1) \Big]\right).
\label{sQ}
\ee
\end{widetext}
Then, it is possible to derive the LQC-modified generalized Friedmann equation:
\be
H^2=\sQ+\frac{\kappa}{3}\rho-\lambda^2\gamma^2\left(\frac{3}{2}\sQ+\frac{\kappa}{3}\rho\right)^2.
\label{fried}
\ee
The details of how to obtain this non-trivial equation are given in the appendix.
It should be pointed out that
\be
\lim_{\lambda\rightarrow 0}\sigma_{Q}^2=\lim_{\lambda\rightarrow 0}\sigma^2,
\ee
so that in the limit $\lambda\rightarrow 0$ the classical Friedmann equation is recovered. On the other hand, in the limit $\sQ\rightarrow 0$, the isotropic holonomy-corrected Friedmann equation is recovered.

From Eq. (\ref{fried}), we can easily find the upper bounds for $\rho$ and $\sQ$:
\be
\rho \leq \rho_c:=\frac{3}{\kappa}\frac{1}{\lambda^2\gamma^2},
\ee
\be
\sQ \leq {\sQ}_c:=\frac{4}{9}\frac{1}{\lambda^2\gamma^2}.
\label{sQc}
\ee

\section{Equations of motion}
\label{EOM}

In the gravitational sector, the information is contained in the combined objects $h_i$:
\be
h_1:=\mub_1c_1=\lambda\sqrt{\frac{p_1}{p_2p_3}}c_1
\qquad \text{and cyclic expressions.}
\ee
It is expected that the six gravitational degrees of freedom $(c_i,p_i)$ account for only three physical degrees of freedom $h_i$. This is because three degrees of freedom are just rescaling of the scale factors which have no physical meaning.

Just as in the classical calculations, we assume isotropic matter. Then we can derive:

\begin{widetext}
\begin{multline}
\doth_1=\frac{1}{N}\{h_1,\ham\}=\frac{1}{2\gamma\lambda}\Big[
(h_2-h_1)(\sin h_1+\sin h_3)\cos h_2 +
(h_3-h_1)(\sin h_1+\sin h_2)\cos h_3\Big] -
\frac{\kappa\gamma\lambda}{2}(\rho+P)
\\ \text{and cyclic expressions,}
\label{doth}
\end{multline}
\end{widetext}
where we have used the constraint $\ham_G+\ham_M=0$. Thus we have
\be
(\doth_i-\doth_j)=-3H(h_i-h_j),
\ee
which means that 
\be
(h_i-h_j)\propto a^{-3} \quad\text{and}\quad \frac{h_i-h_j}{h_i-h_k}=\text{constant}.
\label{hha}
\ee
This should be compared with the classical results given by Eqs. (\ref{interi})-(\ref{interf}).

\section{Symmetries of the effective quantum equations}
\label{sym}
Equations (\ref{h})-(\ref{fried}) are invariant under the discrete  symmetry
\be
\left\{
\begin{array}{l}
h_1 \rightarrow h_1 + (2\tilde{n}_1+\tilde{m}) \pi\\
h_2 \rightarrow h_2 + (2\tilde{n}_2+\tilde{m}) \pi\\
h_3 \rightarrow h_3 + (2\tilde{n}_3+\tilde{m}) \pi
\end{array} \right.
\quad,\quad  
\begin{array}{l}
\forall \tilde{n}_1,\tilde{n}_2,\tilde{n}_3\in \textbf{Z}\\
\forall \tilde{m}\in \{0,1\}.
\end{array}
\label{sym1}
\ee
However, Eq. (\ref{doth}) is only invariant under the smaller symmetry
\be
\left\{
\begin{array}{l}
h_1 \rightarrow h_1 + \tilde{n}\pi\\
h_2 \rightarrow h_2 + \tilde{n}\pi\\
h_3 \rightarrow h_3 + \tilde{n}\pi
\end{array} \right.
\quad,\quad  \forall \tilde{n}\in \textbf{Z}.
\label{sym2}
\ee
Remember that $h_i=\bar{\mu}_i c_i$.

All observable quantities, and their evolution, are invariant under Eq. (\ref{sym2}). This suggests that Eq. (\ref{sym2}) is a gauge symmetry. However, this might not be the case, if more degrees of freedom are taken into account.

More consequences of these symmetries will be discussed later.

\section{Classical limit}

As one would expect, the classical equations are recovered in the limit $\lambda\rightarrow 0$.
But one also expects to find a classical limit in the far future and in the remote past, far away from the bounce. We will therefore investigate for what values of $h_i$ and $\rho$ classical equations are recovered.

For $\sQ\ll{\sQ}_c$ and $\rho\ll\rho_c$, Eq. (\ref{fried}) becomes
\be
H^2=\sQ+\frac{\kappa}{3}\rho,
\ee
to first order in $\sQ$ and $\rho$. The above equation is equivalent to Eq. (\ref{clfried}) if and only if $\sQ=\sigma^2$. It is trivial to check that this is the case, to lowest order in $h_i$ if $h_i\ll 1$. But since $\sQ$ and $\sigma^2$ are cyclic expressions of $h_i$, this is not the only region where Eq. (\ref{clfried}) is recovered from (\ref{fried}).

Eq. (\ref{clfried}) is not enough to completely describe the classical system. To say that we have a classical limit, we also need to recover Eq. (\ref{cleom}). The matter equations are assumed to be unaffected by the holomomy corrections.\\

The symmetries, Eqs (\ref{sym1}) and (\ref{sym2}), suggest the existence of more than one classical limit. And the knowledge of these symmetries could of course be used in the search for such limits.
However, to be absolutely certain that we find all regions of classical behavior, we will search in the full parameter space.

We will try to recover the classical equations, Eqs. (\ref{clfried}) and (\ref{cleom}), from the quantum modified Eqs. (\ref{fried}), in the perturbative regime. But, instead of assuming, for example, that $h_i$ and $\rho$ are small, an thus make an expansion around $(h_i,\rho)=(0,0,0,0)$, we will expand around the more general point $(h_i,\rho)=(h_i^\cl,\rho^\cl)$. We define
\be \label{delta}
\begin{array}{l}
\delta h_i := h_i - h_i^\cl, \\ 
\delta\rho := \rho-\rho^\cl.
\end{array}
\ee

In this section, we will find all points $(h_i^\cl,\rho^\cl)$, such that, for $\delta h_i\ll 1$ and $\delta \rho\ll \rho_c$, Eqs. (\ref{clfried}) and (\ref{cleom}) are recovered from the expressions given in Sections \ref{LQC} and \ref{EOM}.

All the following calculations in this section will be carried out to lowest order in $\delta h_i$ and $\delta \rho$. We will also use the notations
\be \label{Delta}
\begin{array}{l}
\delta\sigma^2 := \sigma^2-\left(\sigma^2\right)^\cl := \sigma^2(h_i)-\sigma^2(h_i^\cl),\\
\delta\sQ\ := \sQ-\left(\sQ\right)^\cl := \sQ(h_i)-\sQ(h_i^\cl).\\
\end{array}
\ee

It should be pointed out at this stage that, even though we assume $\delta\rho\ll\sigma_c$, and indirectly $\delta\sigma^2,\delta\sQ\ll{\delta\sQ}_c$, this does not mean that the energy density and shear have to be small in a classical sens. This is because $\rho_c$ and ${\delta\sQ}_c$ have very large values.\\

Combining Eqs. (\ref{clfried}) and (\ref{fried}) we find that in the classical limit 
\be
\sigma^2 = \sQ -\lambda^2\gamma^2\left(\frac{3}{2}\sQ + \frac{\kappa}{3}\rho\right)^2.
\ee
Expanded, this becomes
\begin{widetext}
\be
\left(\sigma^2\right)^\cl + \delta\sigma^2 = 
\left(\sQ\right)^\cl + \delta\sQ 
-\lambda^2\gamma^2\left(\frac{3}{2}\left(\sQ\right)^\cl + \frac{\kappa}{3}\rho^\cl\right)^2
-2\lambda^2\gamma^2\left(\frac{3}{2}\left(\sQ\right)^\cl + \frac{\kappa}{3}\rho^\cl\right)
\left(\frac{3}{2}\delta\sQ + \frac{\kappa}{3}\delta\rho\right).
\label{expand1}
\ee
\end{widetext}

It should be noticed that $\delta\sigma$ and $\delta\sQ$ are not independent variables since they both depend on $\delta h_i$. However, $\delta\rho$ is independent of $\delta\sigma$ and $\delta\sQ$.
The left-hand side of the above equation does not depend on $\delta\rho$, and since this equation has to be identically fulfilled in the classical limit, the pre-factor in front of $\delta\rho$ on the right-hand side must vanish.
\be
\frac{3}{2}\left(\sQ\right)^\cl + \frac{\kappa}{3}\rho^\cl = 0.
\ee
As $\sQ\geq0$ and $\rho\geq0$ at any time, the only solution is 
\be
\left(\sQ\right)^\cl = \rho^\cl = 0.
\label{zero1}
\ee

Combing the above equation with the definition of $\sQ$ in Eq. (\ref{sQ}), we find:
\be
\cos(h_i^\cl-h_j^\cl)=1,
\ee
which can  be translated into
\be
\begin{array}{l}
h_2^\cl=h_1^\cl + n_2 2\pi \quad,\quad n_2\in\textbf{Z}, \\
h_3^\cl=h_1^\cl + n_3 2\pi \quad,\quad n_3\in\textbf{Z}.
\end{array}
\label{n23}
\ee


The other equation that has to be satisfied in the classical limit is Eq. (\ref{cleom}). 
The left hand side of Eq. (\ref{cleom}) is calculated from $\dot{H}_i=\sum_j \frac{\partial H_i}{\partial h_j}\dot{h}_j$, where $\dot{h}_j$ is given by Eq. (\ref{doth}). The right-hand-side is calculated by inserting expressions for $H_i$ given by Eq. (\ref{h}).

Eq. (\ref{cleom}) should be fulfilled for all $\delta h_i\ll 1$, and therefore also for $\delta h_i=0$. Applying Eq. (\ref{n23}) and $\delta h_i=0$ to Eq. (\ref{cleom}), we get
\begin{widetext}
\be
\frac{\pi(n_2+n_3)}{2\gamma^2\lambda^2}\Big[3-\cos(2h_1^\cl)\Big]\sin(2h_1^\cl)-\cos(2h_1^\cl)\frac{\kappa}{2}(\rho+P) 
=-\frac{\kappa}{2}(\rho+P).
\label{cleom0}
\ee
\end{widetext}
Since this equation has to be identically fulfilled for all matter states, the pre-factor in front of $(\rho+P)$ has to be the same on both sides. Therefore $\cos(2h_1^\cl)=1$, which is equivalent to
\be
h_1^\cl=n_1\pi \quad,\quad n_1\in\textbf{Z}.
\label{n1}
\ee
This also solves the rest of Eq. (\ref{cleom0}).

We also need to recover Eq. (\ref{cleom}) for all $\delta h_i\ll 1$, not equal to zero. Expanding Eq. (\ref{cleom}) to first order in $\delta h_i$ and using Eqs. (\ref{n23}) and (\ref{n1}) we get
\be
\frac{\pi}{\gamma^2\lambda^2}\Big[(n_2+n_3)\delta h_1 + n_3\delta h_2 + n_2\delta h_3\Big]=0.
\ee
For this to be identically fulfilled for all $\delta h_i\ll 1$, we must have $n_2=n_3=0$.
Finally, we find that
\be
h_1^\cl=h_2^\cl=h_3^\cl = n\pi \quad,\quad n\in\textbf{Z}.
\label{cllimit}
\ee

In the classical limit, Eq. (\ref{h}) becomes
\be
H_i=\frac{\delta h_i}{\gamma\lambda}\ll\frac{1}{\gamma\lambda}
\ee
and
\begin{eqnarray}
\sQ&=&\sigma^2,\nonumber\\
&=&\frac{1}{18\gamma^2\lambda^2}\Big[(h_1-h_2)^2+(h_2-h_3)^2+(h_3-h_1)^2\Big],\nonumber\\
&\ll&\sQ_c.
\end{eqnarray}
We also have
\be
\rho=\rho^\cl+\delta\rho=\delta\rho\ll\rho_c.
\ee
This meas that if we are in the classical  limit, the Hubble parameters, the shear and the energy density, are  small compared to the scale of  quantum effects. We want to remind the reader that $\sQ_c$ and $\rho_c$ are of the order of Plank values, which are very large compared to anything expected during most of the evolution of the universe. 

However, $\sQ\ll\sQ_c$ and $\rho\ll\rho_c$ do not guarantee the classical behavior. 
This can bee seen from the symmetries presented in Section \ref{sym}. A change of $h_i$ belonging to the symmetry group Eq. (\ref{sym1}) but not to Eq. (\ref{sym2}) will give unchanged values of $\sQ$ and $\rho$ but will change the dynamics away from the classical one. 
For example, the evolutions in Figs. \ref{unom1} and \ref{unom2} have low energy density through the whole simulation, and passes trough regions of low share and Hubble rates, but never behaves classically.

From the symmetry, Eq. (\ref{sym2}), we can also conclude that all classical limits are equivalent within this framework.
\\

It should be noticed that we have not assumed anything about the pressure. That means that any pressure is allowed in the classical limit.\\

Finally, it should be stressed that this analysis does not claim that the shear cannot be large when compared to the other terms in the classical limit of the Friedmann equations. Usual Bianchi-I can appear as the classical limit of quantum Bianchi-I. Rather, the shear and density have to be small when compared to their maximum allowed values.

\section{Allowed regions in parameter space}

\begin{figure}
	\centering
		\includegraphics[width=1.00\columnwidth]{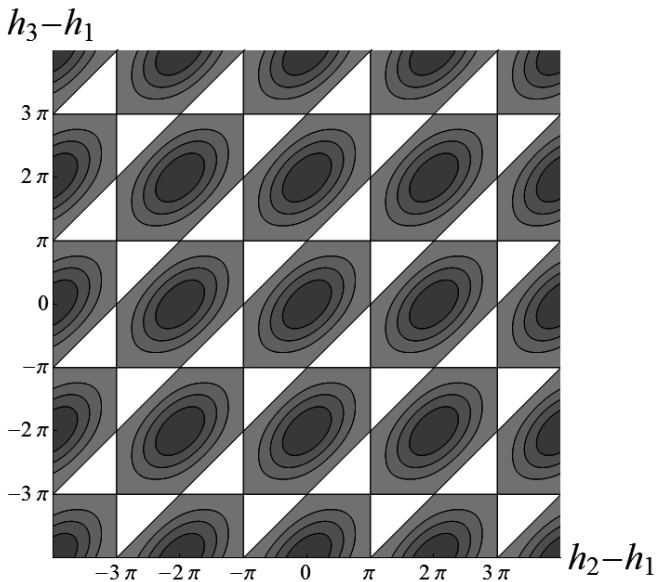}
	\caption{$\sQ$ as a function of $h_2-h_1$ ($x$-axis) and $h_3-h_1$ ($y$-axis). The white areas correspond to $\sQ>{\sQ}_c$, which is forbidden by the modified Friedman equation (\ref{fried}). The black lines are $\sQ=\frac{1}{4}{\sQ}_c,\frac{1}{2}{\sQ}_c,\frac{3}{4}{\sQ}_c,{\sQ}_c$}
	\label{fig:AllowedRegions}
\end{figure}

Fig. \ref{fig:AllowedRegions} displays the parameter space projected down on to $(h_2-h_1,h_3-h_1)$. In this projection, the space is devised into allowed and forbidden regions by the requirement $\sQ\leq{\sQ}_c$. 
The boundaries of those regions, {\it e.g.} when $\sQ={\sQ}_c$, correspond to
\be
h_i-h_j=(2m+1)\pi \quad,\quad i\neq j\ ,\  m\in \textbf{Z}.
\label{boundaryPoints}
\ee

The pattern showed in Fig. \ref{fig:AllowedRegions} goes on infinitely in all directions, which means that there is an infinite number of allowed regions. But, from Eq. (\ref{cllimit}), on can see that there is only one point in this projection near which it is possible to recover the classical limit, and that is $(h_2-h_1,h_3-h_1)=(0,0)$.

An interesting question one can ask is: is it possible, within this framework, to dynamically pass between allowed region? The answers is no, as we shall show in this section.

The allowed regions are only connected by points, therefore any evolution between regions has to pass though these points, defined by:
\be
\begin{array}{l}
  h_j-h_i=(2m_1+1)\pi \\
  h_k-h_i=(2m_2+1)\pi  
\end{array}
\ ,\
\left\{\begin{array}{l}
	i\neq j\neq k\neq i \\
	m_1,m_2\in \textbf{Z}
\end{array}\right. .
\label{connectionpPints}
\ee
Any point on the boundary of the allowed regions, including the points connecting regions can only be reached when $\rho=0$. But even without matter dynamical transitions between regions are impossible. The argument is as follow.

\be
\rho=0\ \Leftrightarrow\ \ham_M=0\ \Leftrightarrow\ \ham_G=0
\ee
which is equivalent to
\be
\sin h_1\sin h_2+\sin h_2\sin h_3+\sin h_3\sin h_1=0.
\ee
Combining the above expression with Eqs. (\ref{connectionpPints}) gives
\begin{widetext}
\be
0=\sin h_i(-\sin h_i)+(-\sin h_i)\sin h_i+(-\sin h_i)(-\sin h_i)
=-\sin^2 h_i.
\ee
\end{widetext}
By once again using Eqs. (\ref{connectionpPints}) with the above relation, one gets:
\be
\begin{array}{l}
	h_i=(m_3-1)\pi\\
  h_j=(2m_1+m_3+1)\pi \\
  h_k=(2m_2+m_3+1)\pi  
\end{array}
\ , \
\left\{\begin{array}{l}
	 i\neq j\neq k\neq i \\
	 m_1,m_2,m_3\in \textbf{Z}.
\end{array}\right. 
\ee
Inserting this into Eq. (\ref{doth}), we obtain $\doth=0$ in all the connection points. Therefore those points can never be dynamically reached. Transitions between the allowed regions displayed in Fig. \ref{fig:AllowedRegions} are not possible, even without matter.

Whatever the region chosen by initial conditions, the solution will stay in that region. In other words, there are infinitely many solutions that never reach a classical limit. However if we assume that the universe starts out in the classical limit of a contracting universe, then the correct region is picked up from the beginning and the evolution will end up in the classical limit of an expanding universe.\\

It is however meaningful to wonder what happened to all the solutions that live in regions without classical limits. We find a clue in Eq. (\ref{hha}).
Since in all the non-classical regions there is a lower bound for at least two of the differences $h_i-h_j$, there must also be an upper bound on $a$. This leaves two possibilities, either the solution approaches a constant $a$ or the solution oscillates forever, leading to multiple bounces. Simulations favor the second hypothesis.

Eq. (\ref{hha}) can be seen as an independent proof of the fact that there is no classical limit in regions not containing $h_i-h_j$ for all $i,j=1,2,3$. Classically, $a$ is unbounded, and this is only possible if $h_i-h_j$ is allowed to be arbitrarily close to zero.

\begin{figure}
	\centering
		\includegraphics{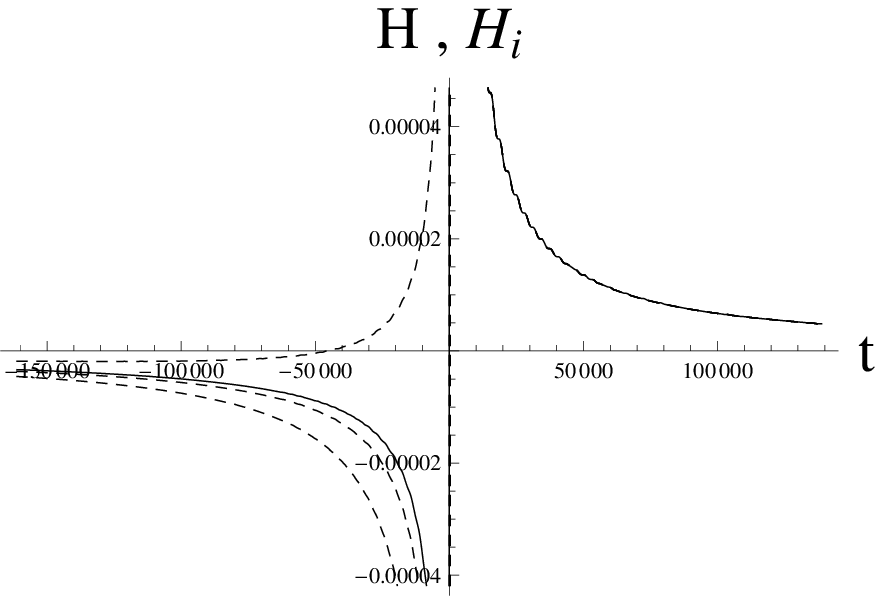}
	\caption{The full line is the total Hubble factor $H$ and the tree dached lines are the directional Hubble factors $H_i$, as a function of time.}
	\label{nom1}
\end{figure}

\begin{figure}
	\centering
		\includegraphics{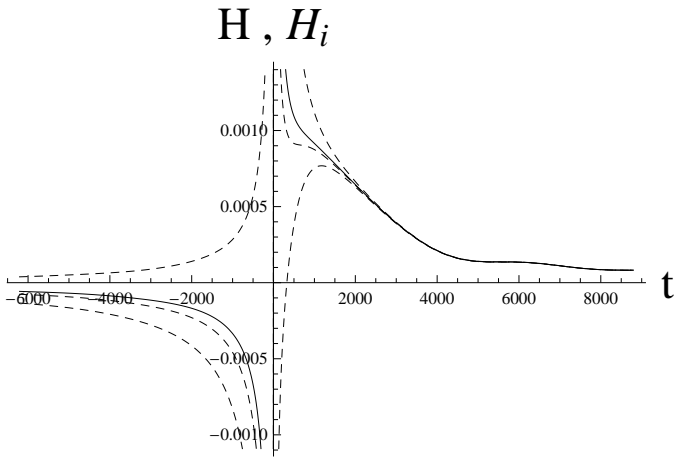}
	\caption{Zoom of Fig. \ref{nom1} round the bounce. The full line is the total Hubble factor $H$ and the tree dached lines are the directional Hubble factors $H_i$, as a function of time.}
	\label{nom2}
\end{figure}

\begin{figure}
	\centering
		\includegraphics{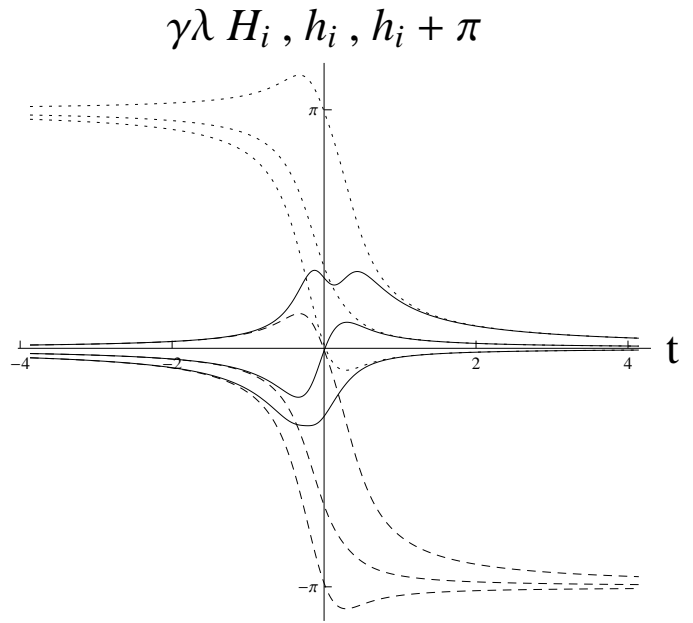}
	\caption{The full lines are $\gamma\lambda H_i$, the dashed lines are $h_i$ and the dotted lines are $h_i+\pi$, as a function of time.}
	\label{nom3}
\end{figure}

\begin{figure}
	\centering
		\includegraphics{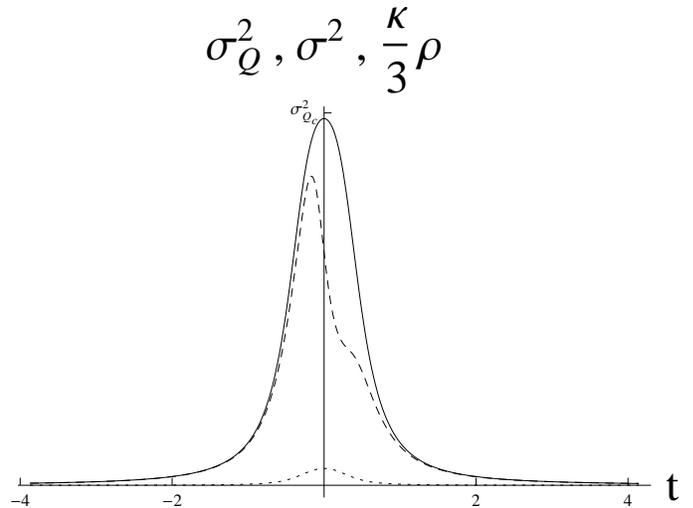}
	\caption{The full line is $\sQ$, the dashed line is $\sigma^2$ and the dotted line is $\rho$, as a function of time.}
	\label{nom4}
\end{figure}

\section{Numerical solutions}
In this section we present some typical examples of numerically generated solutions, both with and without classical limit. In all simulations the matter is taken to be a single massive scalar field, $V(\phi)=m^2\phi^2/2$, $m=10^{-3}$. The equations used in the simulations are Eqs. (\ref{doth}) and the matter equation
\be
\ddot{\phi}+3H\dot{\phi}+m^2\phi=0,
\ee
with $H$ expressed as a function of $h_i$.

Figs. \ref{nom1} - \ref{nom4} are all plots from the same numerical simulations with parameters in the region containing the classical limit. In Fig. \ref{nom1}, we see that, initially, all the directional scale factors are negative but, still in the classical region, one of them changes sign. After the bounce $H_1\approx H_2\approx H_3\approx H$. This is because the matter caused a short inflation -- as can be seen more clearly in Fig. \ref{nom2} which is a zoom around the bounce. Fig. \ref{nom3} is an even closer zoom. Here the quantum effects can be seen. The classical equations are a good approximation until $h_i/(\gamma\lambda)$ deviates from $H_i$. The classical equations become a good approximation again when $H_i\approx (h_i-\pi)/(\gamma\lambda)$. 
During the bounce all the $h_i$ are shifted by $\pi$ compared to $\gamma\lambda H_i$. Simulations suggest that this shift always occurs.
This specific solution exhibits a shear-dominated bounce. This can be seen in Fig. \ref{nom4} since $\sQ\gg\frac{\kappa}{3}\rho$ at the bounce.\\

\begin{figure}
	\centering
		\includegraphics[scale=0.45]{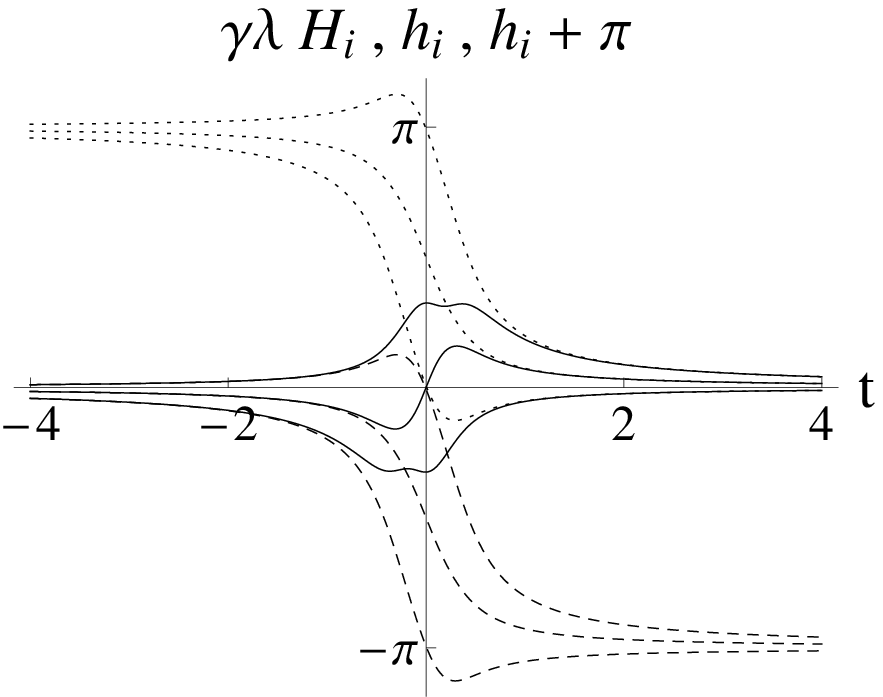}
		\includegraphics[scale=0.45]{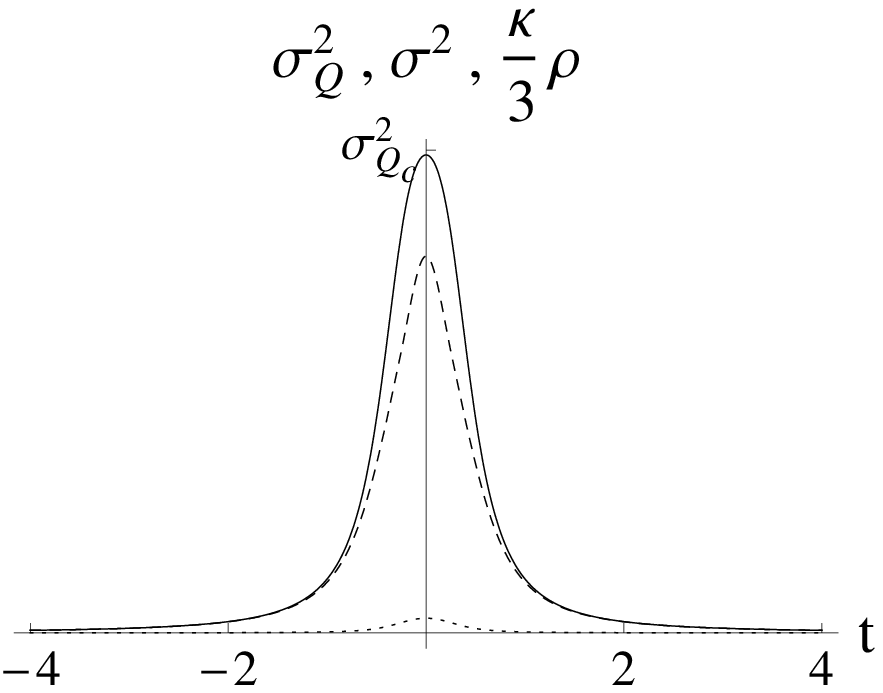}\newline\newline
		\includegraphics[scale=0.45]{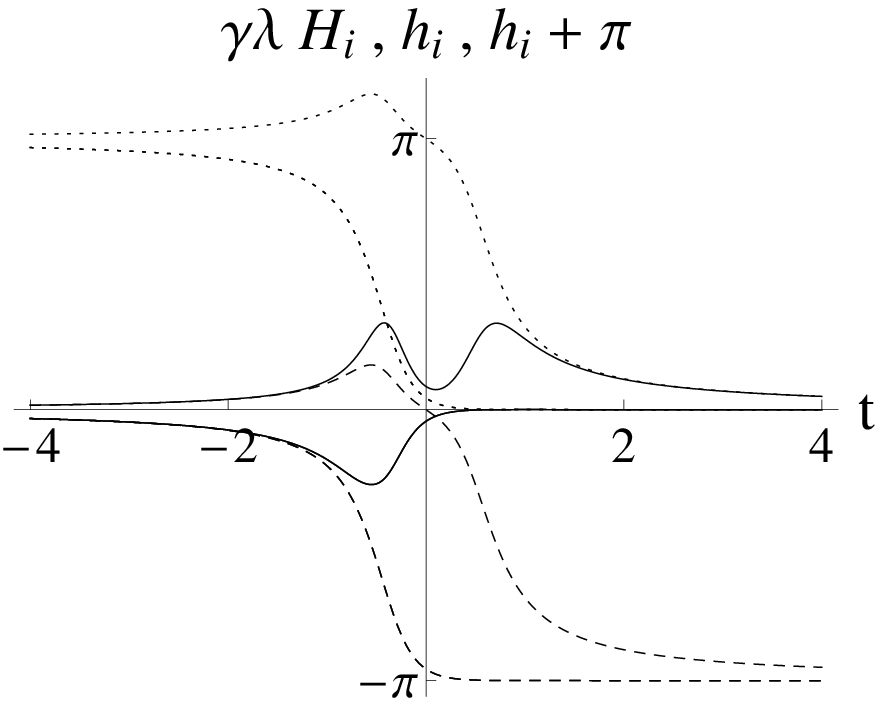}
		\includegraphics[scale=0.45]{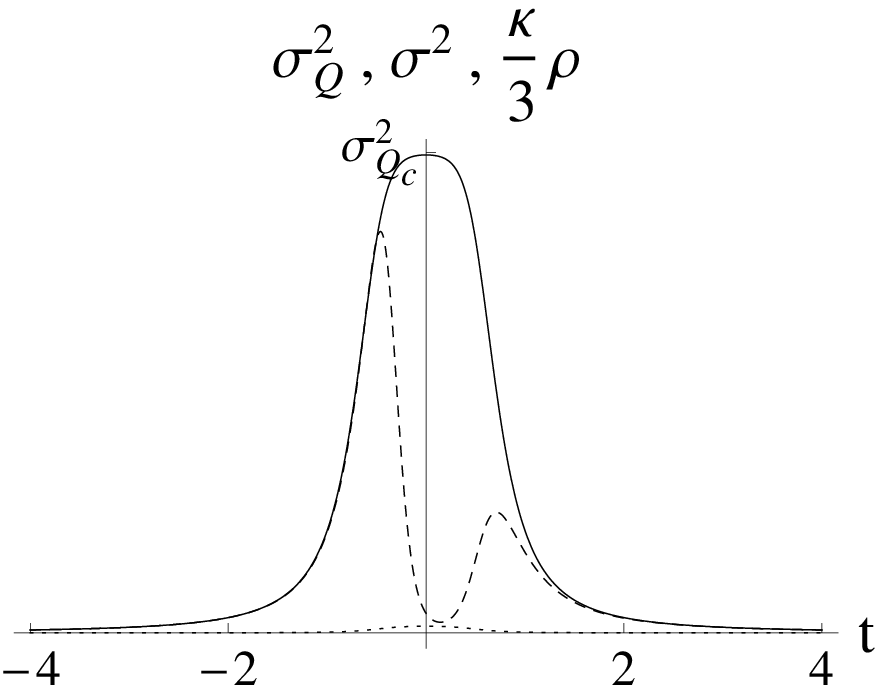}
	\caption{Upper: a solution with maximally symmetric anisotropy. Lower: a solution with  maximally asymmetric anisotropy. Left: the full lines are $\gamma\lambda H_i$, the dashed lines are $h_i$ and the dotted lines are $h_i+\pi$, as a function of time. Right: The full line is $\sQ$, the dashed line is $\sigma^2$ and the dotted line is $\rho$, as a function of time.}
	\label{p4}
\end{figure}

Simulations show that $\sigma^2$ is typically not symmetric around the bounce. For solutions with a classical limit, it appears to be the case that $\sigma^2$ is symmetric around the bounce if and only if $h_i-h_j=h_j-h_k$ for some value of $i\neq j\neq k\neq i$. We therefore call this case maximally symmetric anisotropy. The opposite case is when $h_i=h_j\neq h_k$ for some $i\neq j\neq k\neq i$, and we call this maximally asymmetric anisotropy. Plots similar to Figs. \ref{nom3} and \ref{nom4}, for the maximally symmetric and asymmetric cases are showed in Fig. \ref{p4}.
\\

\begin{figure}
	\centering
		\includegraphics{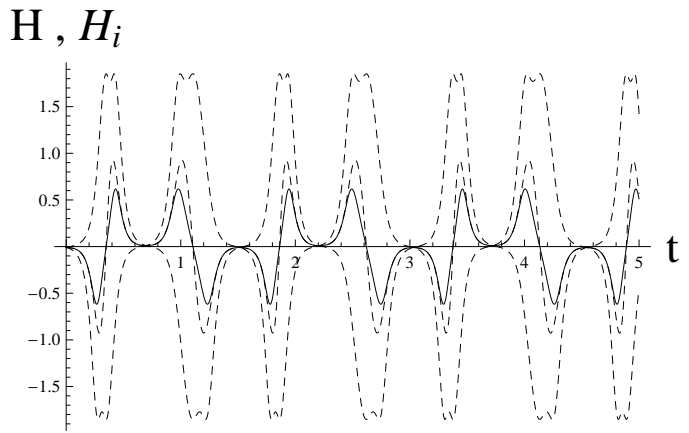}
	\caption{The full line is the total Hubble factor $H$ and the tree dashed lines are the directional Hubble factors $H_i$, as a function of time.}
	\label{unom1}
\end{figure}

\begin{figure}
	\centering
		\includegraphics{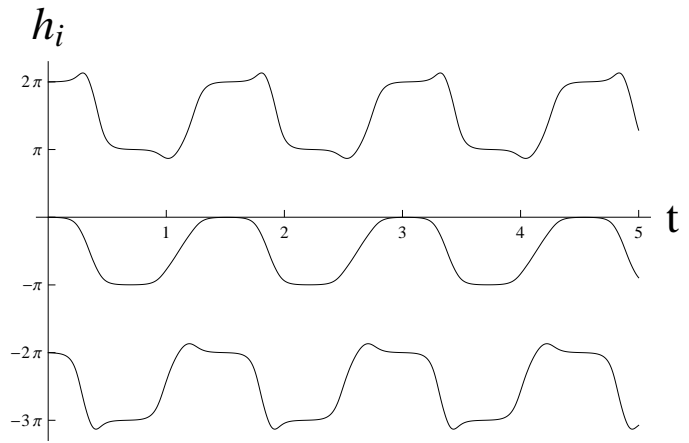}
	\caption{$h_i$ as a function of time in a region without classical limit.}
	\label{unom2}
\end{figure}

Figs. \ref{unom1} and \ref{unom2} are both plots from the same numerical simulations but with parameters in a region with no classical limit. One can see that the behavior is oscillatory and does not resemble anything classically expected.

\begin{figure}
	\centering
		\includegraphics[scale=0.75]{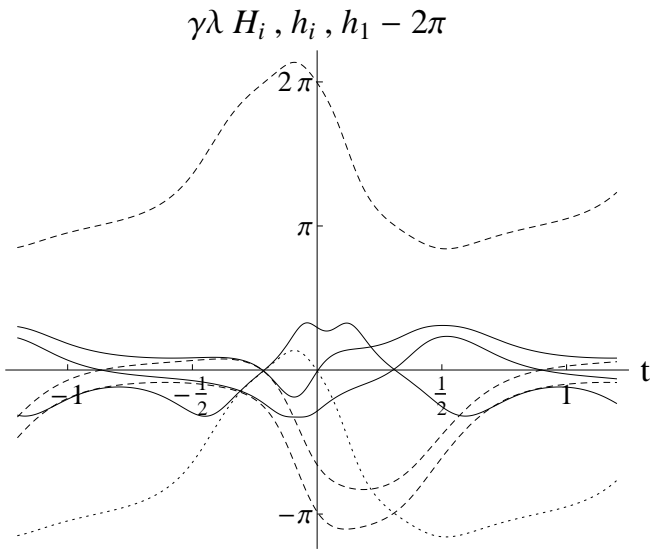}
	\caption{This solution is generated by taking the initial conditions at the bounce with the same values as the solution in Fig. 2-5 but adding $2\pi$ to $h_1$. The full line are $\gamma\lambda H_i$, the dashed lines are $h_i$ and the dotted line is $h_1-2\pi$.}
	\label{r1}
\end{figure}

\begin{figure}
	\centering
		\includegraphics[scale=0.75]{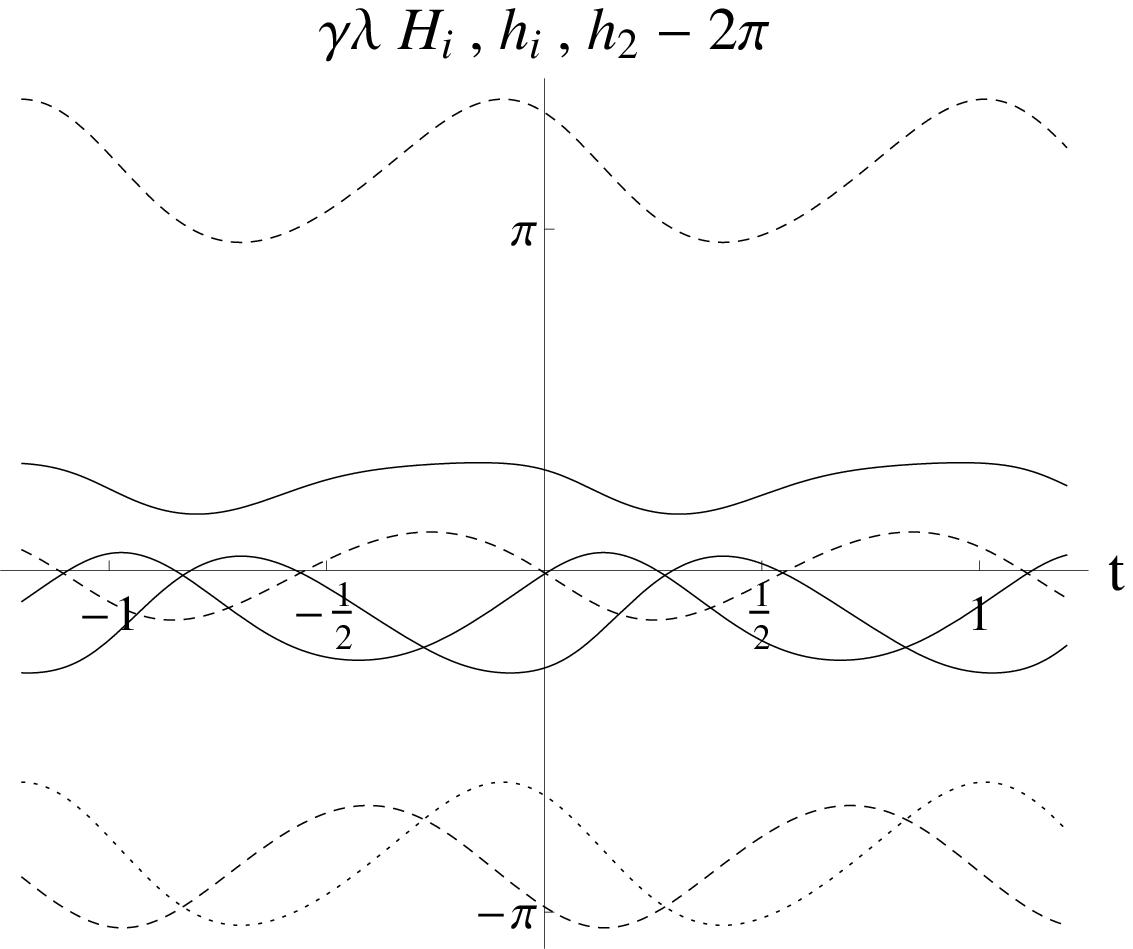}
	\caption{This solution is generated by taking the initial conditions at the bounce with the same values as the solution in Fig. 2-5 but adding $2\pi$ to $h_2$. The full line are $\gamma\lambda H_i$, the dashed lines are $h_i$ and the dotted line is $h_2-2\pi$.}
	\label{r2}
\end{figure}

\begin{figure}
	\centering
		\includegraphics[scale=0.75]{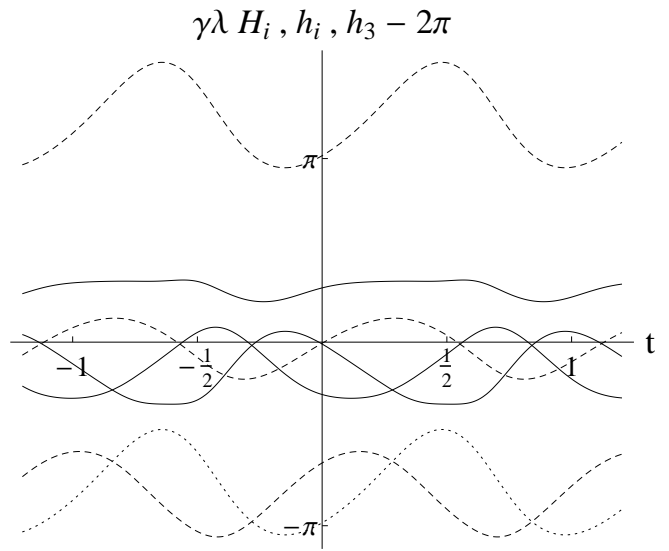}
	\caption{This solution is generated by taking the initial conditions at the bounce with the same values as the solution in Fig. 2-5 but adding $2\pi$ to $h_3$. The full line are $\gamma\lambda H_i$, the dashed lines are $h_i$ and the dotted line is $h_3-2\pi$.}
	\label{r3}
\end{figure}

The simulation shown in Figs. \ref{r1} - \ref{r3} is generated by taking, as initial conditions, the values for $h_i$, $\phi$ and $\dot{\phi}$, as given by the solution shown in Fig. \ref{nom1} - \ref{nom4} at the bounce, with the only difference that $h_i\to h_i+2\pi$ for $i=1,2,3$ respectively. 

Figs. \ref{r1} - \ref{r3} all show oscillatory solutions with periods in the range 1-2 Plank times.
These simulations clearly illustrate that Eq. (\ref{sym1}) is not a symmetry of the full system. There are some similarities between Fig. \ref{nom3} and Fig. \ref{r1}, just around the bounce but the time scale is different by about a factor 4. The solutions in Figs. \ref{r2} and \ref{r3} are very different from anything classical.\\

\section{Discussion}
The results presented in this paper raise an important question for LQC. If the initial conditions are to be put at the bounce, as advocated {\it e.g.} in \cite{abhay}, we face a delicate problem: there are infinitely many more cases leading to universes that do not resemble ours than cases leading to a classically expanding universe. On the other hand, if we set the initial conditions in the classically contracting phase, as advocated in \cite{linda}, we escape this problem. But we face another one: what is the "natural" initial shear? Or, according to which measure --and at which time-- should we assume a flat probability distribution function for variables quantifying the shear? In any case, this requires a deep rethinking of the initial conditions problem.

However, we do not yet know what is the physical meaning of the solutions without classical limit. To understand this better the results presented here, should be compared with, e.g., results found when quantizing this system.

It may also be the case that transitions between different regions in Fig. \ref{fig:AllowedRegions} are possible when including a non zero curvature.\\

This work should also be extended so as to generalize the results presented in \cite{linda}: how will the prediction of the duration of inflation be modified by including anisotropies? This question has been partly addressed already in \cite{Gupt:2013swa}, however, only for a very narrow range of initial conditions.

\acknowledgments
We want to thank the referee for insightful discussions that have led to an important improvement in the formulation of this paper.

This work was supported by the Labex ENIGMASS.

\appendix

\section{Derivation of the modified generalized Friedmann equation}
\label{app}

We define:
\be
s_+=\frac{1}{3}\big[\sin(h_1+h_2)+\sin(h_2+h_3)+\sin(h_3+h_1)\big],
\ee
\be
c_\pm=\frac{1}{3}\big[\cos(h_1\pm h_2)+\cos(h_2\pm h_3)+\cos(h_3\pm h_1)\big].
\ee
The average Hubble parameter can now be written as:
\be
H=\frac{s_+}{2\gamma\lambda}.
\ee
By using elementary trigonometric relations 
$\sin(a)\sin(b)=(\cos(a-b)-\cos(a+b))/2$
and
$\cos(a)\cos(b)=(\cos(a-b)+\cos(a+b))/2$,
we find:
\be
s_+^2+c_+^2=\frac{1+2c_-}{3},
\label{sc}
\ee
and
\be
\ham_G=\frac{3N\sqrt{p_1p_2p_3}}{2\kappa\gamma^2\lambda^2}(c_+-c_-).
\ee
The constraint $\ham_G+\ham_M=0$ then becomes
\be
c_+-c_-=-2\gamma^2\lambda^2\frac{\kappa}{3}\rho.
\label{cc}
\ee
One can now use Eqs. (\ref{sc}) and (\ref{cc}) to rewrite $H^2$ as a function of $c_-$. It will turn out to be useful to expand this expression in terms of $(1-c_-)$. We re-express Eqs. (\ref{sc}) and (\ref{cc}) as
\be
s_+^2=1-\frac{2}{3}(1-c_-)-c_+^2,
\label{ss}
\ee
\be
c_+=1-\left[(1-c_-)+2\gamma^2\lambda^2\frac{\kappa}{3}\rho\right].
\label{c}
\ee
This allows us to write:
\begin{widetext}
\begin{eqnarray}
H^2=\frac{s_+^2}{4\gamma^2\lambda^2} 
&=& \frac{1}{4\gamma^2\lambda^2}\left(1-\frac{2}{3}(1-c_-)-c_+^2\right)\nonumber\\
&=& \frac{1}{4\gamma^2\lambda^2}\left(-\frac{2}{3}(1-c_-)+2\left[(1-c_-)+2\gamma^2\lambda^2\frac{\kappa}{3}\rho\right]-\left[(1-c_-)+2\gamma^2\lambda^2\frac{\kappa}{3}\rho\right]^2 \right)\nonumber\\
&=& \frac{1-c_-}{3\gamma^2\lambda^2}+\frac{\kappa}{3}\rho
-\gamma^2\lambda^2\left(\frac{1-c_-}{2\gamma^2\lambda^2}+\frac{\kappa}{3}\rho\right)^2,
\label{long}
\end{eqnarray}
\end{widetext}
where we have used Eq. (\ref{ss}) for the second equality, and Eq. (\ref{c}) for the third equality.

It can now be seen that, to first order, $(1-c_-)/(3\gamma^2\lambda^2)$ appears just like the shear in the classical Eq. (\ref{clfried}). It can therefore be labeled the quantum shear
\be
\sQ:=\frac{1-c_-}{3\gamma^2\lambda^2},
\ee
which is exactly Eq. (\ref{sQ}). Re-inserting this definition into Eq. (\ref{long}), we find exactly Eq. (\ref{fried}).

\end{document}